\documentclass{moriond}

\def\be{\begin{equation}}
\def\ee{\end{equation}}
\def\bea{\begin{eqnarray}}
\def\eea{\end{eqnarray}}

\newcommand{\Photo}{\includegraphics[height=35mm]{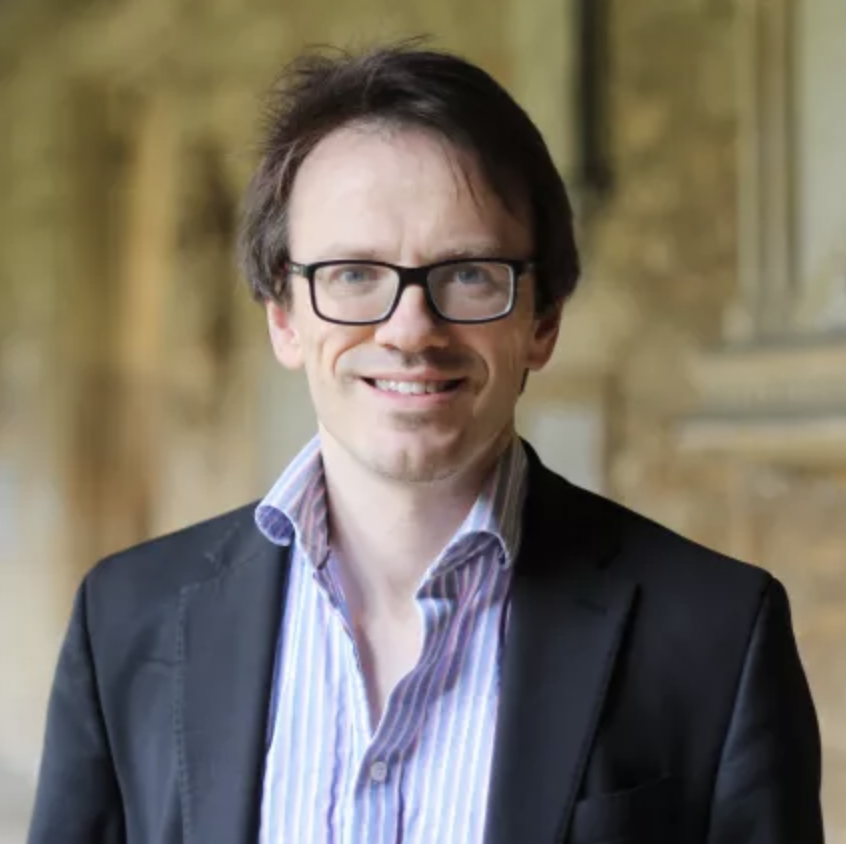}}

\begin{document}
\vspace*{4cm}
\title{String Theory and the Early Universe: Constraints and Opportunities}

\author{ Joseph. P.  Conlon\,\footnote{Photo credit Philippa Sims}}

\address{Beecroft Building, Department of Physics, Parks Road,\\
Oxford OX1 3PU, England}

\maketitle\abstracts{
This is a short account, based on a talk given at the 2024 Moriond Cosmology Conference, of where and why string theory matters in early universe cosmology. It is written for a cosmology audience predisposed to be at best sceptical, and at worst contemptuous, of the notion that either quantum gravity or string theory  has any relevance for their discipline. I cover inflation, CMB tensor modes, extended kination or tracker epochs and reheating.}

\section{Strings and Cosmology }

When does string theory matter to cosmology? When must cosmologists care about string theory?\,\footnote{While I shall use `string theory' throughout this article, most of the arguments are not specific to string theory and could apply to other theories of quantum gravity.} 
If the answer were `never', this would be a short article.

In answering this question, the key organising conceptual idea is that of effective field theory: the idea that theories are good up to a certain energy scale $\Lambda_{UV}$, beyond which extra physics come in, and that low-energy Lagrangians at energies $E \ll \Lambda_{UV}$ can be expanded in
inverse powers of $\Lambda_{UV}$. For the case of string theory (or quantum gravity), the relevant energy scale is the Planck scale, $M_P = 2.4 \times 10^{18} {\rm GeV}$. We encounter this scale (through $G = (8 \pi M_P^2)^{-1}$) in the normal cosmological Lagrangian of general relativity coupled to various matter degrees of freedom
\begin{equation}
\mathcal{L} = \frac{1}{16 \pi G} \int d^4 x \, \sqrt{g} \mathcal{R} + \int d^4 x \, \mathcal{L}_{matter}.
\end{equation}
This GR Lagrangian breaks down at the Planck scale, where new physics (for example, string theory) must come in. However, even assuming string theory to be
 the correct answer to the question of what happens at the Planck scale, how would it manifest itself at much smaller energies, $E \ll M_P$? 

The structure of this question is familiar. It appears in many areas of physics where the ultraviolet theory is known but we are interested in low-energy physics. 
Examples are electroweak and Fermi theory, or the low-energy effective theory of the pions. The well-understood answer is that, within the low energy effective theory, the UV theory manifests itself as non-renormalisable operators suppressed by powers of energy scales associated to the ultraviolet theory. So, in our case the legacy of string theory will be a series of operators expanded with higher powers of $M_P^{-1}$. When can these matter?

Let us start with one example where such operators do \emph{not} matter. Consider a particle accelerator (such as the LHC) operating at a centre of mass energy $E_{CM}$ and studying (for example) Higgs production. Compared to the low energy result, string theory Planck-induced corrections to the cross-section for this process will behave as
\begin{equation}
\sigma = \sigma_0 \left( 1+  \mathcal{O}\left(  \frac{E_{CoM}^2}{M_P^2}  \right) + \mathcal{O}\left(  \frac{E_{CoM}^4}{M_P^4}  \right) + \ldots \right).
\end{equation}
These corrections arise from Planck-suppressed effects; the
extra non-renormalisable terms from string theory make fractional contributions of size $\mathcal{ O}\left(  \frac{E_{CoM}^2}{M_P^2} \right)$ and so are utterly irrelevant to low energy scattering with $E_{CoM} \sim 1 {\rm TeV}$.

The above represents the `standard' intuition that for processes at energies much less than the Planck scales, Planck-scale physics appears via operators which only contribute 
tiny corrections to tree-level physics and so can safely be ignored. In the context of cosmology, the highest energy scales associated to the Standard Cosmology are those during the inflationary epoch, which do not exceed  $E_{inf} \le 10^{16} {\rm GeV}$. Given that even $E_{inf}$ is at least two orders of magnitude lower than the Planck scale, it may then seem that string-scale physics can be safely ignored during any discussion of cosmological evolution, at least in the period during and after inflation. If we do not worry about the interior of black holes and we do not worry about initial cosmological singularities, then string physics is irrelevant and be ignored: such, in brief, is the case for the prosecution.

This argument appears compelling but is flawed. I now describe these flaws and where they show up.

\section{Inflation: Control and Observational Tensors}

The theory of inflation is the leading candidate for the very early universe physics responsible for both the large-scale homogeneities and small-scale inhomogeneities of our universe, through a period of an almost-de Sitter phase lasting approximately sixty efolds. Despite this, there remains much uncertainty as to the right microphysics of the inflationary model. Inflation may be driven by a slowly rolling scalar field $\phi$: but what is that scalar and what was its potential $V_{inf}(\phi)$?

During slow-roll inflation, the potential $V_{inf}(\phi)$ is close to flat. Slow-roll is characterised by the smallness of the two slow-roll parameters, $\epsilon, \eta \ll 1$ where $\epsilon$ and $\eta$ are defined as
\begin{equation}
\epsilon = \frac{M_P^2}{2} \left( \frac{ V'(\phi)}{V(\phi)} \right)^2,
\end{equation}
\begin{equation}
\eta = M_P^2 \frac{V''(\phi)}{V(\phi)}.
\end{equation}
A nearly scale-invariant power spectrum requires $\epsilon, \eta \ll 1$. Once either $\epsilon$ or $\eta$ becomes $\mathcal{O}(1)$, then inflation terminates rapidly (and also can no longer generate the close-to-scale-invariant spectrum of perturbations observed in the CMB).

If we are happy to treat inflation simply as a paradigm, then we could stop here. However, much work on inflation has gone into constructing explicit inflationary models in order to provide causal explanations \emph{why} the universe underwent an extended de Sitter phase and \emph{why} this resulted in a close-to-scale-invariant spectrum.

This is the point at which string theory, in the guise of Planck-suppressed operators, enters the picture. As stressed above, at energies $E \ll M_P$ the footprint of string theory 
takes the form of Planck-suppressed operators. String theory matters when such operators matter. To investigate this, suppose we have an inflationary potential $V_{inf}(\phi)$ which we modify through addition of additional Planck suppressed operators quadratic in the inflaton,
\begin{equation}
V_{inf}(\phi) \to V_{inf}(\phi) + \frac{\phi^2}{M_P^2}  V_{inf}(\phi)
\end{equation}
The additional term is a Planck suppressed contribution to the potential. Provided that the field $\phi$ only moves through sub-Planckian distances, it has only a small effect on the value of $V(\phi)$. However, it is apparent that in terms of the slow-roll eta parameter $\eta$, this term results in
\begin{equation}
\eta \to \eta + 2,
\end{equation}
and so this Planck-suppressed operator contributes an $\mathcal{O}(1)$ correction to the $\eta$ parameter, destroying the validity of the inflationary model. Although first formulated in supergravity and sometimes called the `supergravity $\eta$-problem', it is clear from the above argument that the issue is general and not tied to supergravity in any way.

A causal explanation of why an inflationary model can generate a long period of slow roll inflations therefore requires a good argument why this dangerous $\frac{\phi^2}{M_P^2} V_{inf}(\phi)$ operator is absent. As such a term is Planck-suppressed, any such argument necessarily requires an understanding of Planck-suppressed operators and thus Planck scale physics.\footnote{Note that simply appealing to `symmetries' is \emph{not} a good enough argument to avoid having to think about the Planck scale, not least because global symmetries are generally broken by Planck-scale physics (`quantum gravity has no global symmetries').} There are various approaches to controlling this term.\cite{Baumann:2014nda} The two main solutions are those of (a) approximate symmetries where symmetry-breaking is small enough to suppress the coefficient of this operator and (b) an accidental cancellation in which the operator coefficient is not protected but happens to be small in a particular model.

The $\eta$-problem relates to the way Planck-suppressed operators affect control over the \emph{curvature} of the potential. Another question, particularly relevant to models involving large tensor modes in the CMB, involves control over the potential itself. During inflation, the inflaton $\phi$ moves in field space. This field displacement $\Delta \phi$ can be related to 
the tensor-to-scalar ratio $r$ that is a target for many CMB experiments.\cite{Baumann:2014nda}
\begin{equation}
\label{Lyth}
\frac{\Delta \phi}{M_P} > \left( \frac{r}{0.01} \right)^{1/2}.
\end{equation}
The significance of Eq. \ref{Lyth} is that the observability of tensor modes is correlated with the field displacement; in particular, observable tensor modes require field displacements $\Delta \phi$ that are close to $M_P$.

This brings in one uncontroversial and one slight more contested fact to do with string theory and inflation. The uncontroversial fact is that justifying a slow-roll inflationary epoch involving a trans-Planckian field excursion requires justifying the flatness of the potential over this transPlanckian distance in field space. In the general modification of a potential by Planck-suppressed operators,
\begin{equation}
V(\phi) \to V(\phi) \left( 1 + \alpha \frac{\phi}{M_P} + \beta \frac{\phi^2}{M_P^2} + \gamma \frac{\phi^3}{M_P^3} + \dots \right)
\end{equation}
it is clear that any justification of flatness over transPlanckian distances requires control over \emph{all} the Planck-suppressed operators (note that this is a much stronger condition than the $\eta$ problem, which only requires control over the quadratic term in the above expansion). Any such control can only be achieved through the ultraviolet theory, namely string theory. Inflationary models leading to large tensors, if they exist, can therefore only be \emph{understood} in a quantum gravity context: the low-energy theory lacks the tools to control such operators.

The other, more contested, area is whether it is \emph{ever} possible in string theory to have potentials that are flat over trans-Planckian distances. The term `trans-Planckian' is unfortunately a bit loose here: in effective field theory arguments, it is hard to argue over the privileged status of $M_P$ compared to $\frac{M_P}{\pi}$, $\frac{M_P}{2}$ or $2 \pi M_P$. The sharpest statements apply for $\Delta \phi \gg M_P$ where such $\mathcal{O}(1)$ factors do not matter.

In such a parametric limit, it appears impossible for string theory potentials to remain flat over such distances. Despite a lot of effort, no examples are known. Furthermore, in explicit studies of field ranges, symmetries that can protect potentials (for example, the axion shift symmetry) are only good for sub-Planckian (or, at best, $\mathcal{O}(1)$ Planckian) distances. Field ranges which can easily be trans-Planckian involve fields such as the volume modulus or dilaton that are directly involved in the string scale\,\footnote{when formulating the physics in 4d Einstein frame with a fixed value of $M_P$ (for example, $M_s = M_P/\sqrt{\mathcal{V}}$).} and so cannot have flat potentials over long distances. This lack-of-flatness relates to the ideas of the swampland distance conjecture\,\cite{Ooguri:2018wrx}, the statement that towers of states become exponentially light over transPlanckian field displacements, $m_i \sim M_0 \exp \left( - \lambda \Delta \Phi / M_P \right)$: the light tower of states reaches scales comparable to Hubble and then back-react on the effective field theory, violating flatness. 

Indeed, it is perhaps the case that the current failure to observe tensor modes in the CMB is a necessary feature of quantum gravity and that large-field inflation models of the sort that would have produced observable tensor modes are incompatible with quantum gravity.

Turning things around, the above two arguments can be viewed as a clue to why inflationary model-building (as opposed to the inflationary paradigm) has been frustrating: successful inflation models require control over Planck scale physics.

\section{Kination and Tracker Epochs}

What happens after inflation ends is largely unknown. In the standard picture of cosmological evolution, the period of inflation is followed by rapid reheating and a conversion of energy into relativistic degrees of freedom which redshift as radiation and generate the Hot Big Bang. However, there are currently minimal observational constraints on the era between inflation and nucleosynthesis\,\cite{Allahverdi:2020bys}, leaving plenty of opportunity for qualitatively different cosmological behaviours. Examples are \emph{kination} or \emph{tracker} epochs. In the former, the universe is dominated by the kinetic energy density of a rolling scalar field whereas in the latter there are fixed proportions of kinetic energy, potential energy and a background fluid (e.g. radiation) as a field rolls down an exponential potential. Such epochs are also well motivated in string constructions where the final vacuum may be at a large distance in field space from the inflationary locus and have been the focus of much recent work in 
string compactifications\,\cite{Apers:2022cyl} \cite{Shiu:2023fhb} \cite{Revello:2023hro} \cite{Apers:2024ffe} \cite{Seo:2024qzf}.

Such epochs are also considered in the regular cosmology literature. Why is string theory (or some theory of the Planck scale) necessary to understand and justify them?

During a kination epoch, in which the potential can be neglected, it follows from the evolution equations
\begin{eqnarray}
\ddot{\Phi} + 3 H \dot{\Phi} & = & 0, \\
3 H^2 M_P^2 & = & \frac{\dot{\Phi}^2}{2},
\end{eqnarray}
that the kinating scalar $\Phi$ evolves as
\begin{equation}
\Phi(t) = \Phi(t_0) +  \sqrt{\frac{2}{3}} M_P \ln \left( \frac{t}{t_0} \right)
\end{equation}
with $a(t) \sim t^{1/3}$. This evolution implies that the scalar traverses approximately one Planckian distance in field space every Hubble time. In any extended kination epoch lasting multiple Hubble times, whether in the early or late universe, the kinating field traverses mutiple Planckian distances in field space. 

One of the easiest ways to realise a kination epoch is through a scalar field rolling down a steep exponential potential $V(\Phi) = V_0 e^{- \lambda \Phi / M_P}$. If the potential is steep enough ($\lambda > \sqrt{6}$) then the field enters a kination epoch as the potential energy grows ever more sub-dominant. However, as energy densities redshift during a kination epoch as
\begin{equation}
\rho_{kin} \sim \frac{1}{a(t)^6},
\end{equation}
any other fluid (in particular, radiation or matter) will grow in importance relative to the kinating scalar and eventually catch up. For exponential potentials, the endpoint of this evolution is a \emph{tracker} solution with fixed proportions of the energy density in potential, kinetic and fluid energy.

During a tracker solution along an exponential potential $V(\Phi) = V_0 e^{- \lambda \Phi / M_P}$, the evolution of the scalar field is
\begin{equation}
\Phi(t) = \Phi(t_0) + \frac{2 M_P}{\lambda}  \ln \left( \frac{t}{t_0} \right).
\end{equation}
As the tracker solution only exists for $\lambda \ge \sqrt{6}$, we see that the scalar field evolution is always
slower than in kination epochs; nonetheless, for reasonable values of $\lambda$ it remains true that the field $\Phi$ traverses approximately one Planckian distance in field space every Hubble time while traversing a significantly transPlanckian distance in field space during any extended tracker epoch lasting multiple Hubble times.

The relevance of string theory is now clear: the consequence of the above is that it is not possible to incorporate either extended kination or extended tracker epochs into a cosmological history without reckoning with Planck scale physics in the form of Planck-suppressed operators. When a field traverses a transPlanckian distance $\Delta \Phi \gg M_P$ in field space, any Lagrangian becomes vulnerable to corrections of the form $f \left( \frac{\Delta \Phi}{M_P} \right)$
which could disrupt this evolution: and so neither kination nor tracker epochs can be justified and understood by themselves without control over such Planck scale physics.

In the context of string theory, there exist more precise statements about field behaviour over such trans-Planckian distances in field space, in particular through the \emph{Swampland Distance Conjecture}\,\cite{Ooguri:2018wrx} (which by now has a large amount of support from many explicit examples). This states that such field excursions are accompanied by a tower of states becoming exponentially light, so that the masses of this tower of states behaves as
\begin{equation}
m_i \sim M_0 e^{-\lambda (\Delta \Phi) / M_P},
\end{equation}
where $\lambda$ is an $\mathcal{O}(1)$ constant. As such states become exponentially light, the effective UV cutoff of the original effective field theory becomes correspondingly lower. One well-understood example of such a tower of states is the Kaluza-Klein tower present in string compactifications as fields evolve towards the large volume decompactification limit.
Any theory operating with a fixed UV cutoff then becomes invalid as this tower of states descends and goes below the cutoff.

Here it is useful to clarify one misconception that sometimes appears. The Swampland Distance Conjecture is not a statement that effective field theories break down or cannot be used as fields evolve through trans-Planckian field displacements (such a statement would be incorrect). Rather, it is a statement about towers of states becoming light; a statement about the behaviour of the UV theory in this limit. There are well-established cases where the UV theory is known and effective field theories continue to be valid for arbitrarily transPlanckian field displacements. The clearest example of this is the decompactification limit: in the large radius limit (which is at a substantially trans-Planckian distance in field space from the small-volume region), the effective field of string compactifications is well understood and becomes better (not worse) controlled. The fact that this might exist at 20 $M_P$ from the centre of moduli space is neither here or there: the control parameter is $\frac{l_s}{R}$. The only subtlety is that the \emph{cutoff} of the theory (which is tied to the KK scale) also becomes parametrically small as fields evolve to large volume.\footnote{This is somewhat analogous to the behaviour of quantum field theories as a function of renormalisation scale. For example, the use of the QCD Lagrangian defined at 100 GeV to compute scattering at $10^{14}$ GeV results in large logarithms $\ln \left(10^{14} / 10^2 \right)$ appearing. These logarithms do not mean that the theory breaks down or becomes incomputable at $10^{14}$ GeV: instead, we should simply work with the renormalised couplings that are defined at $10^{14}$ GeV. Likewise, in the large radius limit of string compactifications we should regard the UV cutoff as a radius-dependent quantity and treat this as a dynamical quantity; doing so, the 4d effective theory becomes progressively better controlled as the internal radius becomes larger.}

\section{Reheating}

The endpoint of kination or tracker epochs will generally be a vacuum solution, with fields oscillating about the vacuum. Many of the most interesting string theory vacua are in the asymptotic regions of moduli space. Examples are the Large Volume Scenario\cite{Balasubramanian:2005zx} or the DGKT construction\cite{DeWolfe:2005uu}. If these are the true vacuum of the theory, fields have to evolve on paths ending on them. This motivates the study of extended kination/tracker epochs that can reach such vacua starting from the interior of moduli space.
However, independent of the path towards the final minimum, at some point there needs to be a process of \emph{reheating} to transfer energy into relativistic Standard Model degrees of freedom.

The Planck scale is an unavoidable shadow haunting the physics of reheating, always important whether a theorist likes it or not. It cannot simply be wished away. Why?
The true, if apparently facetious, reason lies in one of the important equations of cosmology,
\begin{equation}
\label{eq34}
3 < 4 .
\end{equation}
The centrality of Eq. \ref{eq34} in cosmology is that the energy density in radiation behaves as $\rho_{\gamma} \sim a^{-4}$ whereas the energy density in matter behaves as $\rho_{m} \sim a^{-3}$. Matter that hangs around, and does not decay, inevitably comes to dominate over radiation. In reheating, the race is not to the swift, but the slow: as energy is transferred from matter to radiation, decay products of early decaying particles redshift away rapidly and it is the \emph{last} particles to decay, not the first, that dominate the physics of reheating.

And this is why the Planck scale is always relevant to reheating: on dimensional grounds, the decay rate of a typical Planck-coupled 
modulus scalar $\Phi$ is \cite{Banks:1993en} \cite{deCarlos:1993wie}
\begin{equation}
\Gamma_{\Phi} \sim \frac{\lambda}{8 \pi} \frac{m_{\Phi}^3}{M_P^2},
\end{equation}
where $\lambda$ is an $\mathcal{O}(1)$ numerical constant. $M_P^{-1}$-suppressed interactions in a Lagrangian result in $M_P^{-2}$ factors as amplitudes
are turned into physical decay rates: the numerator factor of $m_{\Phi}^3$ then follows on dimensional grounds. 
In contrast, a particle $\psi$ which has renormalisable decay modes will have a decay rate
\begin{equation}
\Gamma_{\psi} \sim \frac{\lambda}{8 \pi} m_{\psi},
\end{equation}
which is enhanced by a factor $\left( \frac{M_P}{m_{\psi}} \right)^2$ compared to particles that can only decay via Planck-suppressed modes.

This decay rate for moduli then implies a lifetime
\begin{equation}
\tau_{\Phi} \sim \frac{8 \pi}{\lambda} \frac{M_P^2}{m_{\Phi}^3}  \sim \frac{1}{\lambda} \left( \frac{10^6 {\rm GeV} }{m_{\Phi}} \right)^3 \sim 10^{-6} {\rm s}.
\end{equation}
A modulus of mass $m_{\Phi} \sim 10^6 {\rm GeV}$ (for example) will live a long time and, almost inevitably, come to dominate the energy density of the universe. 
The relative redshifting of matter and radiation, combined with the long lifetime of moduli, implies that even if moduli are present with tiny initial abundances we expect them, over time, to come to dominate the energy density of the universe. 

Note that this physics of decays, where matter dominates over radiation, inverts the usual logic of scattering processes, where the weakest interactions are the \emph{least} important. 
In terms of reheating, it is those (massive) particles that interact most weakly that are the \emph{most} important. Particles that interact more strongly decay more rapidly; their relativistic decay products redshift as radiation and rapidly become sub-dominant to any non-relativistic matter that remains. The Planck scale is the largest scale in physics and gravity is the weakest force: particles which interact only via Planck-suppressed interactions have the longest lifetimes (relative to their mass).

We do not currently know whether there exist particles with only Planck-suppressed interactions. We do know of particles which only interact via the electromagnetic force (the right-handed electron at low energies), particles which only interact via the weak force (the neutrinos) and particles which only interact via the strong force (the gluon). It would, therefore, not be surprising if there did exist scalars or fermions which only interact via gravitational-strength interactions. Based on the nomenclature for such particles in string theory, we refer to these as moduli.

It is precisely because the weak inherit the early universe that any understanding of reheating requires an understanding of the spectrum of moduli. A modulus with mass 1000 TeV, coupling only via Planck-suppressed interactions, is utterly irrelevant for particle physics. However, it will survive until only shortly before the era of Big Bang Nucleosynthesis, come to dominate the energy density of the university even if originally present with low abundance, and reheat the universe to a relatively low temperature, parametrically given by
\begin{equation}
T_{reheat} \sim \frac{m_{\Phi}^{3/2}}{M_P^{1/2}} \sim \left( \frac{m_{\Phi}}{10^6 {\rm GeV}} \right)^{3/2} 1 \, {\rm GeV} .
\end{equation}
Such low reheat temperatures have important consequences for the physics of dark matter production and baryogenesis, as various models for these rely on much larger initial reheat temperatures.

When reheating is driven by Planck-coupled moduli, the decay modes of such moduli are also crucial to the physics of reheating. There is no especial reason why Planck-suppressed interactions should favour the Standard Model sector over other kinematically accessible sectors in a compactification. In particular, if light relativistic degrees of freedom (such as axions or axion-like particles) exist, then moduli can decay to these as $\Phi \to aa$ and these can in principle give rise to dangerous amounts of dark radiation.\,\cite{Cicoli:2012aq}\,\cite{Higaki:2012ar}
\section{The Universe Today}

We would be lucky indeed if string theory is directly relevant to the cosmology of the present universe. There is, however, one clear way this could happen. If a network of cosmic superstrings\,\cite{Copeland:2009ga} exist -- and if the tension $\mu$ satisfies $G \mu < 10^{-7}$ then this is not excluded -- then the universe, today, could be filled with a network of fundamental strings. This is not impossible; but it is also something of a lottery ticket and so, while we mention on it, we do not dwell on it.

 \section{Conclusions}
 
 String theory matters when the Planck scale matters. In this article, I have argued that, contrary to naive expectation, there are many places in the early universe which are sensitive to Planck-suppressed physics. The physics of quantum gravity is then not some esoteric topic only relevant to cosmologists unfortunate enough to fall through black hole horizons; it also forcibly intrudes on those who want to build and control inflationary models, study the implications of extended kination and/or tracker epochs and understand the physics of reheating.

In summary: even if you are uninterested in string theory or quantum gravity, it is interested in you and your theories. In those darkest of dark ages in the time before nucleosynthesis, it skulks like a wraith of the night to overturn and destroy those cosmological models and histories that pay it insufficient respect.

\section*{Acknowledgments}

This article is based upon work from COST Action COSMIC WISPers CA21106, supported by COST (European Cooperation in Science and Technology).
I acknowledge support from STFC consolidated grants ST/T000864/1 and ST/X000761/1. I thank the organisers of the 2024 Moriond Cosmology conference for the invitation to give the talk on which this article is based.

\end{document}